\documentclass[12pt]{article}

\usepackage{chngcntr}
\usepackage{scicite}
\usepackage{graphicx}
\usepackage{times}
\usepackage{multirow}
\usepackage[braket]{qcircuit}
\usepackage{xcolor}

\usepackage{lineno}

\newenvironment{sciabstract}{%
\begin{quote} \bf}
{\end{quote}}

\title{Observation of measurement-induced quantum phases in a trapped-ion quantum computer}

\author{Crystal Noel,$^{1,3,4,\ast}$ Pradeep Niroula,$^{1,2}$
Daiwei Zhu,$^{1}$ Andrew Risinger,$^{1}$ \\
Laird Egan,$^{1}$ Debopriyo Biswas,$^{1}$ Marko Cetina,$^{1,3}$ Alexey V. Gorshkov,$^{1,2}$ \\ Michael J. Gullans,$^{2}$ David A. Huse,$^{5}$ Christopher Monroe$^{1,2,3,4,6}$ \\
\normalsize{$^{1}$Joint Quantum Institute, Departments of Physics and Electrical and Computer Engineering}
\\ \normalsize{NIST/University of Maryland, College Park, MD 20742}\\
\normalsize{$^{2}$Joint Center for Quantum Information and Computer Science,}\\
\normalsize{NIST/University of Maryland, College Park, MD 20742}\\
\normalsize{$^{3}$Duke Quantum Center and Department of Physics, Duke University, Durham, NC 27708}\\
\normalsize{$^{4}$Department of Electrical and Computer Engineering, Duke University, Durham, NC 27708}\\
\normalsize{$^{5}$Department of Physics, Princeton University, Princeton, NJ 08540}\\
\normalsize{$^{6}$IonQ, Inc., College Park, MD  20740}\\
\normalsize{$^\ast$To whom correspondence should be addressed; E-mail:  crystal.noel@duke.edu.}
}

\date{}

\begin{document} 


\baselineskip24pt


\maketitle



\begin{sciabstract}
 Many-body open quantum systems balance internal dynamics against decoherence from interactions with an environment. Here, we explore this balance via random quantum circuits implemented on a trapped-ion quantum computer, where the system evolution is represented by unitary gates with interspersed projective measurements. As the measurement rate is varied, a purification phase transition is predicted to emerge at a critical point akin to a fault-tolerant threshold. We probe the ``pure'' phase, where the system is rapidly projected to a deterministic state conditioned on the measurement outcomes, and the ``mixed'' or ``coding'' phase, where the initial state becomes partially encoded into a quantum error correcting codespace.  We find evidence of the two phases and show numerically that, with modest system scaling, critical properties of the transition emerge.
 
\end{sciabstract}

An isolated many-body quantum system undergoes unitary evolution until it is probed by its environment via quantum measurement \cite{CarmichaelBook,GardinerBook}.  The irreversible process of measurement converts quantum coherence in the system into classical entropy in the measurement apparatus due to the intrinsic randomness of quantum measurements.  When the rate of partial measurements is high, this process ``collapses'' the many-body system into a pure quantum state consisting of locally correlated regions determined by the recent unitary dynamics and measurement outcomes.  At low measurement rates, however, there is a mixed (coding) phase where the associated projections can leave invariant a codespace in the system that retains  memory of initial conditions for exponentially long times \cite{Skinner19,Li18,Gullans20,Choi20}.  Such measurement-induced phase transitions have recently been theoretically explored in models based on random quantum circuits, but are believed to be a ubiquitous phenomenon in monitored non-equilibrium quantum systems.  The theory of these transitions, although still nascent, has seemingly deep connections to percolation and conformal field theory \cite{Skinner19,Li19,Jian20,Bao20}, as well as threshold theorems in fault-tolerant quantum computing \cite{Aharonov00}.  Observing these effects in experiment is a formidable challenge because measuring the observables that signify the transition requires exquisite control and isolation of the system, accurate monitoring by an external measurement apparatus, and the use of sophisticated feedback or post-processing with the measurement data. 

Here, we report on a direct experimental observation of the two phases associated with a measurement-induced purification transition in a trapped ion quantum computer. We use a single reference qubit initially entangled with the system to directly test for the existence of the codespace in the mixed phase and its absence in the pure phase  \cite{Gullans20b}.  This approach has the practical benefit that it relaxes experimental resource requirements compared to observables that require measuring entanglement entropies of large numbers of qubits, such as measuring Renyi entropy \cite{Brydges2019}. We avoid the use of  post-selection on measurement outcomes through the addition of feedback operations that reverse any measurement-induced unitary rotations on the reference qubit (i.e., so-called ``quantum steering'' effects \cite{Schrodinger36}). As a result, absent noise, our experimental approach is directly scalable to large systems.   

From early measurements of the quantum-to-classical nature of measurement in ion trap systems \cite{WinelandNobelRMP} and cavity quantum electrodynamics \cite{HarocheNobelRMP}, to the recent observation of wavefunction collapse in superconducting qubits \cite{jump2019}, the phenomenon of measurement itself has been a subject of great interest experimentally. 
Many-body coherent operations combined with controlled dissipation or measurements have been explored experimentally in, for example, the study of dissipative state preparation \cite{Barreiro11}, as well as in recent theoretical proposals for many-body quantum non-demolition measurements \cite{Yang20}. 
We also note related experimental results showing symmetry resolved dynamical purification of spin chains in a long-range $XX$ model with local depolarizing noise \cite{vitale2021symmetryresolved, Brydges2019}. By contrast, in our study, we employ a ``digital'' model of computing with two-site unitaries and projective measurements with a temporal randomness to the dynamics.  

Our quantum computer uses up to 13 $^{171}{\rm Yb}^+$ qubits in a single chain of 15 trapped ions in a microfabricated chip trap~\cite{Maunz2016}. We achieve native single-qubit gate fidelities of 99.96\% and two-qubit gate fidelities on any pair of 98.5-99.3\%, as detailed elsewhere~\cite{egan2021faulttolerant}.

We now describe the specific dynamics of the random circuits in this work with a system of $L$ qubits subject to unitary evolution with all-to-all connectivity and measurements.      For such all-to-all coupled models, spatial entanglement of the wavefunction is not a reliable diagnosis of the measurement-induced phase transition; instead we characterize the problem in terms of a purification transition  \cite{Gullans20}.  In this picture, the system transitions  at low measurement rates to a phase with long-range correlations in time, similar to the behavior found in   fault-tolerant error correction thresholds. This dynamical purification phase transition  can be efficiently probed by studying how the system preserves entanglement over time with a single reference qubit \cite{Gullans20b}.

An example circuit is shown in Fig.~1A. After preparing all qubits in $\ket{0}$, the reference is entangled to a randomly selected system qubit to form a Bell pair. The entangling operation is followed by a scrambling unitary, which consists of random single-qubit Clifford gates and two-qubit $XX(\pi/4)$ gates on random qubit-pairs. The scrambling stage (Supplementary Materials), spreads the entanglement to the entire system and reduces finite-size effects. After scrambling the system qubits (Supplementary Materials), we evolve the system in time with random unitary dynamics and measurements with a total number $N_g = \lfloor L \sqrt{L}\rfloor$ of $XX(\pi/4)$ gates applied to randomly chosen qubit pairs.

After each of the entangling-gates, with probability $p$, we add a measurement (Supplementary Materials). While mid-circuit readout of ion qubits is possible \cite{fossfeig2021entanglement}, we instead use extra qubits available in the system as ancilla qubits to defer readout to the end of the circuit. 
When a circuit calls for measurement on a system qubit, it is entangled with an ancilla after rotation to a chosen measurement basis. Because the unitaries are XX gates, the measurement choice of the $z$ or $x$ basis has a strong effect on the subsequent dynamics. This feature of our model
allows us to tune the probability, $p_x$, that a measurement is in the $x$ basis to go across the purification transition without having to directly change the measurement probability $p$. At the end of the circuit, all the qubits, including the system qubits, reference qubit and measurement ancillae, are read out in the $z$-basis via fluorescence imaging. For each circuit, we rotate the reference qubit to measure in $x$, $y$ and $z$-basis and post-select the observations to obtain Pauli expectations conditioned on measurement outcomes (Supplementary Materials). The set of three Pauli expectations for each measurement outcome are then used to construct the density matrix of the reference qubit and measure its entropy $S_Q$. 
These circuits are examples of stabilizer circuits, whose noiseless dynamics are efficiently simulable on a classical computer through the Gottesman-Knill theorem \cite{Gottesman98,Aaronson04}. 

 As an illustrative example, in Fig.~1B, we consider the experimentally measured evolution of the reference qubit entropy in two circuits sampled from ensembles with $p=0.15$. One circuit sampled from $p_x=0$ stays mixed (encoded) and one with $p_x=1$ purifies over time. Units of time are measured in number of applied two-qubit gates, $N_g$, for consistency between theory and experiment.
 For noiseless stabilizer circuits, the entropy is always either 0 or 1 bit \cite{Aaronson04,Li19}, and, as a result, the circuits that purify must do so at precisely one time step. However, this property no longer holds exactly in the presence of noise.
 Experimentally, we find that the mixed circuit maintains a high value of the entropy of the reference qubit $S_Q$. In the second circuit, the reference qubit purifies at the expected time in the circuit, albeit to a constant offset in entropy due to experimental noise. It is already apparent from these examples that we observe a clear separation between pure and mixed results for the entropy of the reference qubit. For each of these circuits, we ran 4000 shots of each measurement basis ($x,y,z$) for each circuit to compute the entropy of the reference qubit at each time step. 

In order to characterize the many-body dynamics of the circuits in this model, we generate large ensembles of circuits and average their entropy for given values of $p$, $p_x$, and $L$. In Fig.~2A, we show the theoretical phase diagram for the model vs $p$ and $p_x$.  For low $p$ and $p_x$, the system is driven to a mixed (coding) phase where the non-unitary dynamics projects quantum information about the initial state into a random quantum error correcting code. As either $p$ or $p_x$ is increased, the  system enters a pure phase, where an initial mixed state collapses to a fixed quantum state and the encoding operation fails.
The behavior at $p=0$ can be smoothly connected to the finite $p$ behavior by scaling time  by $1/p$, in which case the purification dynamics is described by a measurement-only model \cite{Ippoliti21}.  By this convention, the system can be in a pure phase even for infinitesimal values of $p$.   The critical point at each value of $p$ was obtained from finite-size scaling analysis using simulations of $L=16$ to $L=64$ qubits (Supplementary Materials).  Our scaling analysis is based on extracting the exponential decay rate of $\langle S_Q(t) \rangle$ at late times.

In Fig.~2B, we show the simulated dynamics of $\langle S_Q(t) \rangle$ at two representative points in the phase diagram with $p=0.15$. 
In the mixed phase, probed at $p_x=0$, $\langle S_Q \rangle$ stays near one for exponentially long times in $L$ and serves as a local order parameter for the phase.  Deep in the pure phase where $p_x= 1$, the reference qubit rapidly purifies, with an average entropy that exponentially approaches zero. In the experiment, we probe small systems $L\le 8$ after a number $L^{1.5}$ of gate operations.  For larger numbers of qubits $L$, this scaling limit is sufficient to probe the phase because the effective depth of the circuit scales as $2\sqrt{L}$, much greater than any fixed correlation time in the system. At the critical point, as we show in the Supplementary Materials (Fig.~\ref{fig:decaytime}), the entropy decay time scales as $L^{1/5}$ to conform to the universal critical dynamics of the system.

In order to reduce the number of circuits needed to evaluate the order parameter, we append a feedback circuit to the end of each circuit that is expected to purify. The feedback uses single-qubit rotations and a Boolean logic circuit of CNOT gates between the reference and measurement ancillae to return the reference to the zero state in the $z$-basis, thus disentangling it from the measurement ancillae (details in Supplementary Materials). With this addition, we replace measurement of the quantum entropy $S_Q$ with the classical one $S_C$, and eliminate the need to also measure in the $x$-basis and $y$-basis. 
The use of such a feedback circuit also avoids the need to post-select on measurement outcomes.
The extension of this feedback approach, which relies on classical ``decoding'' of the measurement outcomes \cite{Gullans20b}, beyond stabilizer circuits in the large-$L$ limit,  has potentially interesting connections to computational complexity theory.

To probe the phases experimentally, we generate an ensemble of random circuits for the chosen values of $p$, $p_x$, and $L$ to run on the ion trap quantum computer. To constrain the number of measurements to a low value, we study a fixed line of parameters at $p = 0.15$ (Fig.~2A), and the evolution is applied for a time $N_g$.
At the end of the circuit, we measure the reference in the $z$-basis. We average over many shots to determine the classical entropy, $S_C$, of each circuit. The majority of experimental noise can be explained with a simple noise model using XX-gate crosstalk. (See Supplementary Materials, where we also describe techniques to further mitigate errors).
We assume a Gaussian distribution of expected $S_C=0$ circuit outcomes and $S_C=1$ circuit outcomes and find their intersection, which is used as a threshold at $S_C=0.93$ (Fig.~\ref{fig:supp_histograms}, Fig.~\ref{fig:supp_rawdata}). Any outcome below the threshold is counted as $S_{C,T}=0$, and those above as $S_{C,T}=1$. 
 For $p_x=0,1$ ($p_x=0.5)$, we average the entropy after binning with the threshold, $\langle S_{C,T} \rangle$, over the results of 300 (100) unique circuits. 

We study $\langle S_{C,T} \rangle$ at $p_x = (0,0.5,1)$ and $L=(4,6,8)$, and observe the first experimental evidence of the phases of a dynamical purification phase transition. We find that while the measured entropy increases with system size in the mixed phase ($p_x=0$), the opposite behavior is observed in the pure phase ($p_x=1$) as the entropy decreases with system size (Fig.~3A). This behavior is expected and can readily be seen in simulations at the experimental probe time in the example in Fig.~2B. To probe the crossover behavior on these system sizes, we also sample at an intermediate value of $p_x=0.5$ close (for these sizes) to the critical point at $p_{xc} = 0.72(1)$. 
We observe consistent results with the simulations in this near-critical regime, showing behavior that interpolates between the two extremes.

Having obtained conclusive evidence for the two phases in our system, it remains an outstanding challenge to experimentally probe the universal critical behavior of this model. We predict that such effects will become accessible in our system through modest increases in system sizes from $L = 8$ to $L=32$ qubits combined with periodic sympathetic cooling \cite{cetina2020quantum}, which enables mid-circuit measurements and should allow for deep circuits. We have found that a sensitive probe of the critical properties of the purification transition is the late-time exponential decay constant $\tau$ of the order parameter $\langle S_Q(t)\rangle \sim e^{-t/\tau}$.  Fig.~3B shows an example of a finite-size scaling analysis that can be used to extract critical properties of the model.  Here, we use direct simulations of the ideal circuit evolution to predict the behavior of our system as it is scaled to larger sizes.  Crucially, these scaling results illustrate that the critical properties of the purification transition are obtainable using the modest systems sizes and circuit depths accessible in near-term ion-trap hardware. 

Although more delicate than conventional equilibrium phases of matter, our results show that measurement-induced quantum phases are accessible in near-term quantum computing systems despite the formidable experimental challenges.  Recent years have seen a host of advances in mapping out the phenomenology of these novel nonequilibrium phases of matter, including the prediction of  topological order stabilized by measurements in random circuits \cite{Lavasani21,Sang20,Ippoliti21} and applications in computational complexity theory \cite{Napp19} and quantum error correction \cite{Gullans20c}. These developments point to a broad potential for the advancement of many-body physics and quantum information science through the continued explorations of quantum measurement.

From the perspective of fault-tolerant quantum computation, our results open up a number of new directions. An important conceptual aspect of our work is that we experimentally study an error correction threshold as a physical phenomena, exploring its connections to universality in quantum many-body physics.  This approach contrasts with many prior works on experimental quantum error correction that have so-far focused primarily on demonstrations of few-body gadgets in the below-threshold regime \cite{egan2021faulttolerant}. In addition, while error correction thresholds can be studied numerically, applying those theories in practice requires a deeper understanding of the errors in real physical systems, and how the corresponding thresholds behave. The ability to successfully operate quantum computing systems in these near-critical regimes is likely to be a crucial aspect in the future of fault-tolerant quantum computing. Therefore, our work also represents an important demonstration in quantum error correction and fault-tolerance, made possible by the combined experimental and theoretical advances described in this work.

\begin{figure}
\begin{centering}
\includegraphics[scale=.88]{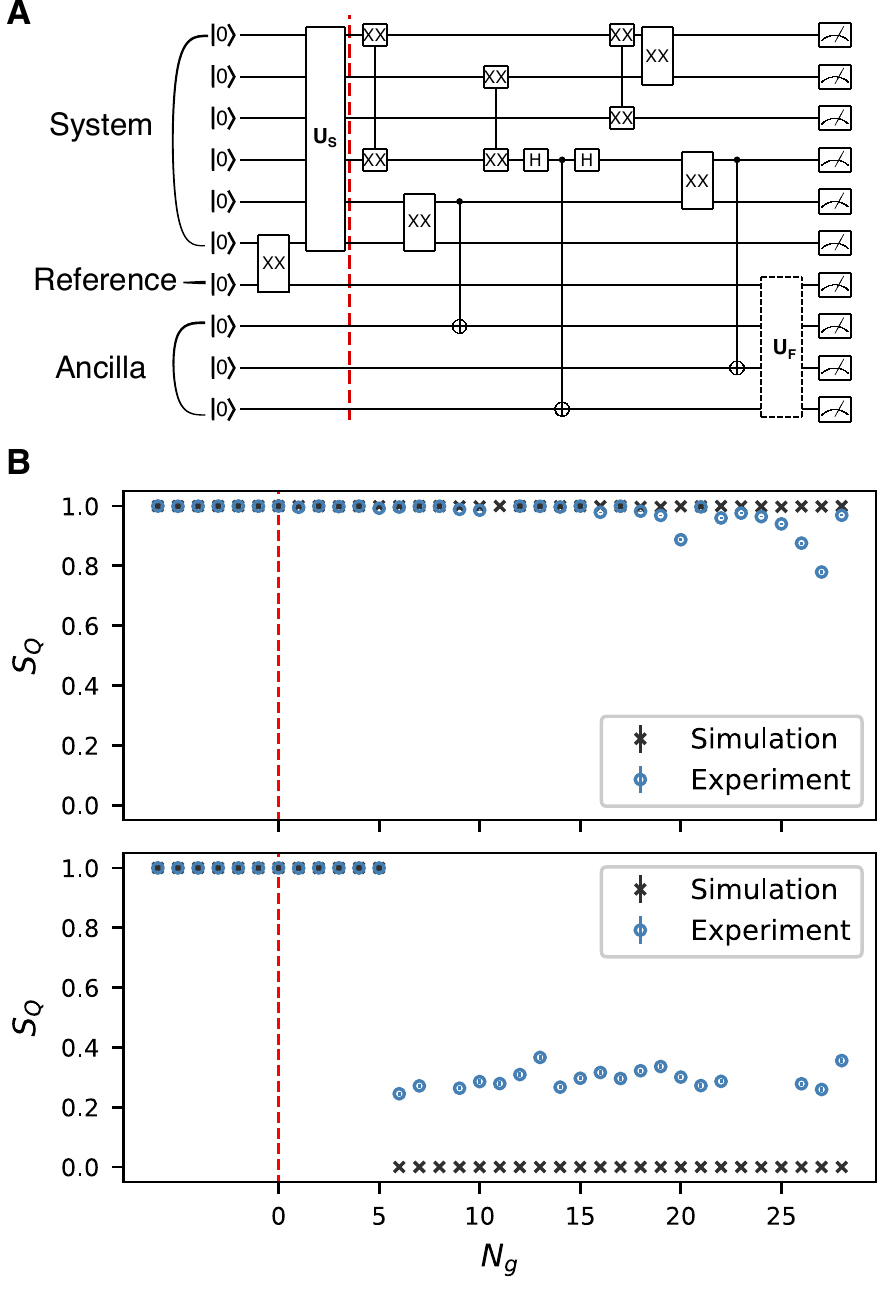}
\caption{\label{fig:fig1} \textbf{Model and Purification Dynamics} (A) Schematic of a circuit with $L=6$ system qubits, $N_g=6$ two-qubit gates, 2 $z$-measurements and 1 $x$-measurement. The first XX gate entangles the reference with a system qubit. Next, we scramble the system, $U_S$. The time evolution of the unitary-measurement dynamics starts at the red dashed line. Probabilistic measurement is deferred until the end of the circuit using CNOT gates between system qubits and measurement ancillae. The x-basis measurement is shown after the third $XX$ gate. Finally, a feedback operation $U_F$ is applied (see Supplementary Materials) (B) The entropy of reference qubit for two $L=6$ circuits where the reference qubit stays mixed (upper panel) and purifies (lower panel). The $x$-axis shows the evolution of time in units of applied two-qubit gates ($N_g$) after scrambling is complete (indicated again by the red dashed line). In this example, the entropy is measured by performing single-qubit tomography of the reference by making measurements in the $x$, $y$ and $z$-basis. Error bars ($1\sigma$) are smaller than the markers, with 4000 and 10000 shots for experiment and simulation, respectively. Missing experiment data are due to ion loss events, which are assumed to be uncorrelated with the data being taken. }
\end{centering}
\end{figure}

\begin{figure}
\begin{centering}
\includegraphics[scale=0.88]{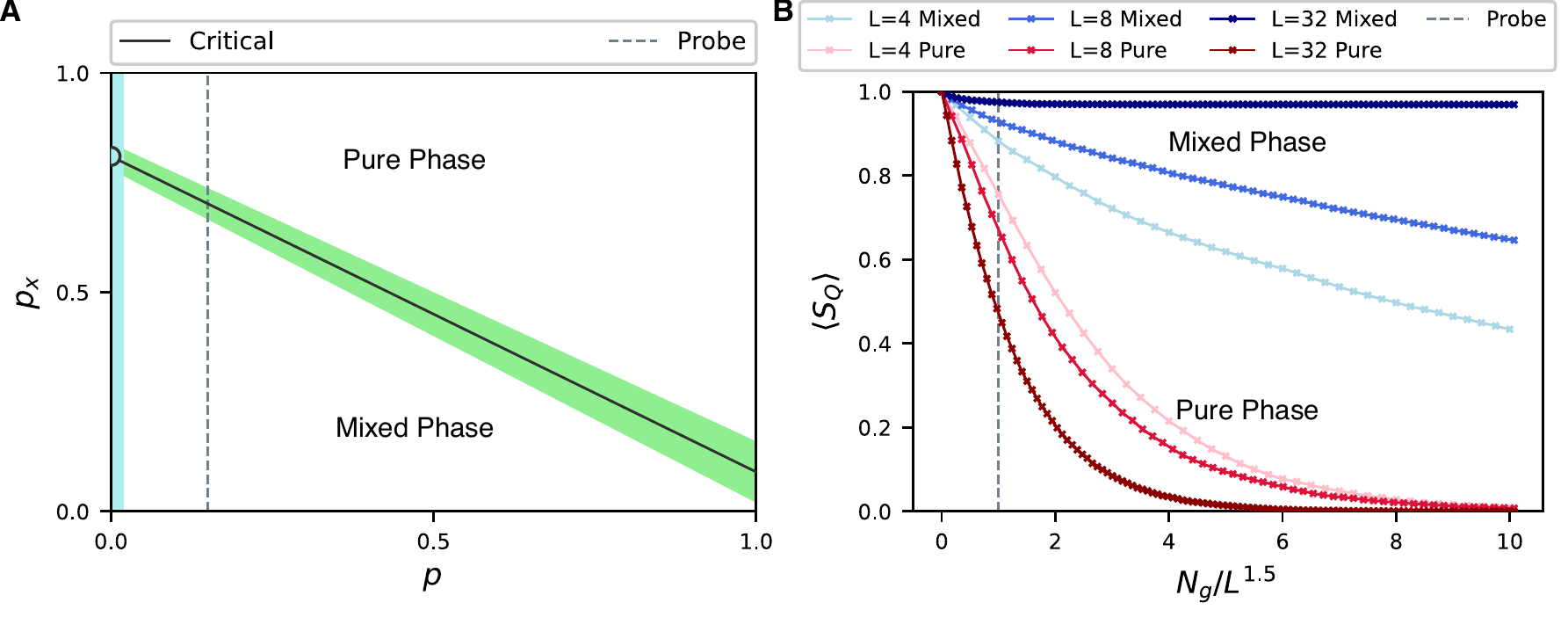}
\caption{\label{fig:diagram} \textbf{Phase Diagram and Scaling Limit of Average Purification Dynamics} (A) The phase diagram of the model, parameterized by $p$ and $p_x$. The green shaded region indicates the numerical uncertainty in the critical region between the top phase where the reference qubit rapidly purifies, and the bottom phase where it stays mixed. In our experiment, we fix $p=0.15$, and tune $p_x$ to probe the phase transition along the dashed line.  In the limit $p\to 0$ (left blue shaded region) with time also scaled as $1/p$, our model becomes a measurement-only model \cite{Ippoliti21}. A purification transition (circle) arises in this limit when tuning $p_x$ along the line $p=0^+.$
(B) The simulated entropy of the reference qubit averaged over many random circuits $\langle S_Q\rangle$ in the two phases. Here, we use the same fixed value of $p=0.15$ from A, with $p_x=0$ (mixed) and $p_x=1$ (pure) plotted against time (measured in units of two-qubit gates) scaled by $L^{1.5}$. The dashed vertical line indicates the experimental probe time of $N_g = L^{1.5}$, and the intersection of this line with different system sizes shows increasing (decreasing) entropy in the mixed (pure) phase that is the signature of the two phases.}
\end{centering}
\end{figure}

\begin{figure}
\includegraphics[scale=0.88]{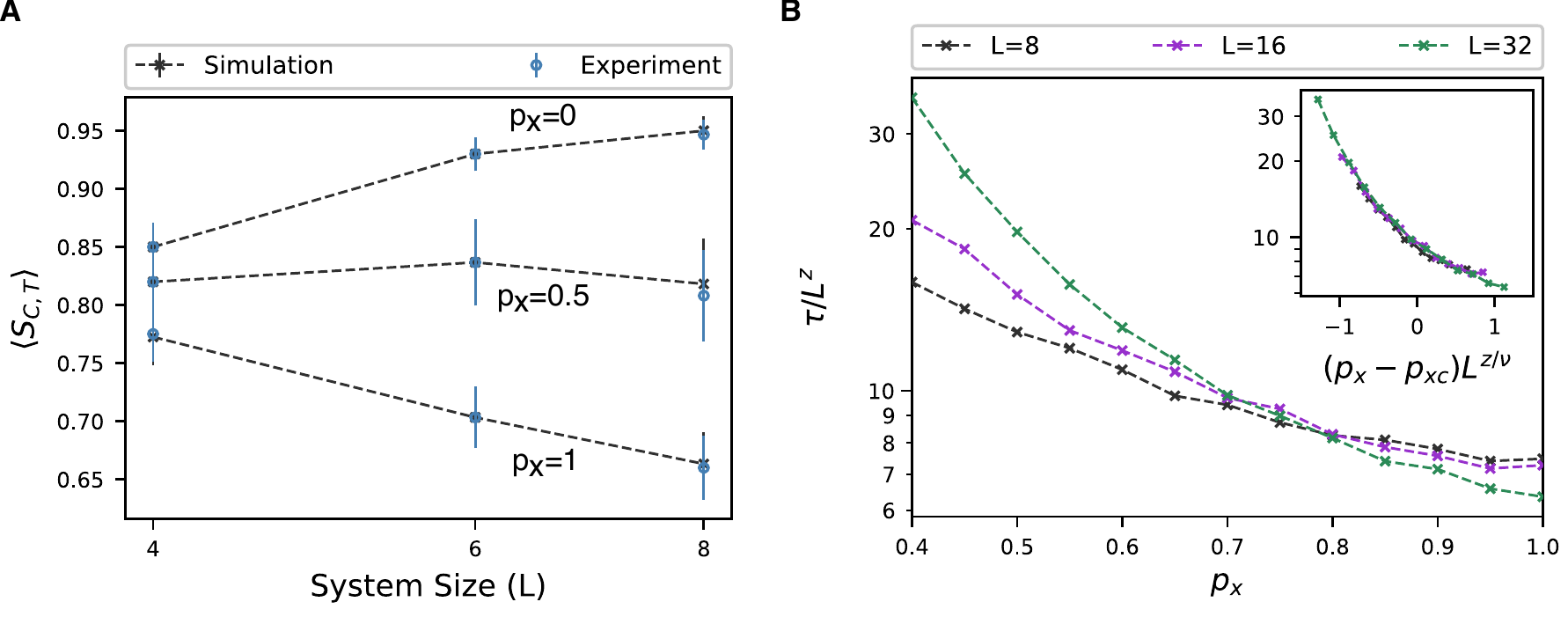}
\caption{\textbf{Experimental Observation of Phases and Simulated Critical Behavior} (A) Classical entropy after thresholding $\langle S_{C,T} \rangle$ averaged over an ensemble of random circuits at varying system sizes. We show evidence of mixed (top), intermediate (middle), and pure phase (bottom) with $p_x=0$, $p_x=0.5$, and $p_x=1$ respectively with size-scaling as predicted in Fig.~2B. Error bars are $1\sigma$ uncertainty with 300 circuits for $p_x=0,1$ and 100 circuits for $p_x=0.5$ (B) Simulated results showing the late-time decay rate $\tau$ of $\langle S_Q \rangle$ near the transition.  Here, $z\approx 1/5$ is the dynamical critical exponent, $\nu \approx 1/2$ is the correlation length exponent, and $p_{xc}=0.72(1)$ is the critical value of $p_x$. These critical parameters are extracted from a finite-size scaling analysis (see Inset and Supplementary Materials).  \label{fig:results}}
\end{figure}

\clearpage

\bibliography{micbib.bib}

\providecommand{\noopsort}[1]{}\providecommand{\singleletter}[1]{#1}%
\begin{thebibliography}{10}
\expandafter\ifx\csname url\endcsname\relax
  \def\url#1{\texttt{#1}}\fi
\expandafter\ifx\csname urlprefix\endcsname\relax\def\urlprefix{URL }\fi
\providecommand{\bibinfo}[2]{#2}
\providecommand{\eprint}[2][]{\url{#2}}

\bibitem{CarmichaelBook}
\bibinfo{author}{Carmichael, H.}
\newblock \emph{\bibinfo{title}{An open systems approach to quantum optics}}
  (\bibinfo{publisher}{Springer}, \bibinfo{address}{Berlin, Germany},
  \bibinfo{year}{1993}).

\bibitem{GardinerBook}
\bibinfo{author}{Gardiner, C.~W.} \& \bibinfo{author}{Zoller, P.}
\newblock \emph{\bibinfo{title}{Quantum Noise}} (\bibinfo{publisher}{Springer},
  \bibinfo{address}{Berlin, Germany}, \bibinfo{year}{2000}).

\bibitem{Skinner19}
\bibinfo{author}{Skinner, B.}, \bibinfo{author}{Ruhman, J.} \&
  \bibinfo{author}{Nahum, A.}
\newblock \bibinfo{title}{Measurement-induced phase transitions in the dynamics
  of entanglement}.
\newblock \emph{\bibinfo{journal}{Phys. Rev. X}} \textbf{\bibinfo{volume}{9}},
  \bibinfo{pages}{031009} (\bibinfo{year}{2019}).
\newblock \urlprefix\url{https://link.aps.org/doi/10.1103/PhysRevX.9.031009}.

\bibitem{Li18}
\bibinfo{author}{Li, Y.}, \bibinfo{author}{Chen, X.} \&
  \bibinfo{author}{Fisher, M. P.~A.}
\newblock \bibinfo{title}{Quantum zeno effect and the many-body entanglement
  transition}.
\newblock \emph{\bibinfo{journal}{Phys. Rev. B}} \textbf{\bibinfo{volume}{98}},
  \bibinfo{pages}{205136} (\bibinfo{year}{2018}).
\newblock \urlprefix\url{https://link.aps.org/doi/10.1103/PhysRevB.98.205136}.

\bibitem{Gullans20}
\bibinfo{author}{Gullans, M.~J.} \& \bibinfo{author}{Huse, D.~A.}
\newblock \bibinfo{title}{Dynamical purification phase transition induced by
  quantum measurements}.
\newblock \emph{\bibinfo{journal}{Phys. Rev. X}} \textbf{\bibinfo{volume}{10}},
  \bibinfo{pages}{041020} (\bibinfo{year}{2020}).
\newblock \urlprefix\url{https://link.aps.org/doi/10.1103/PhysRevX.10.041020}.

\bibitem{Choi20}
\bibinfo{author}{Choi, S.}, \bibinfo{author}{Bao, Y.}, \bibinfo{author}{Qi,
  X.-L.} \& \bibinfo{author}{Altman, E.}
\newblock \bibinfo{title}{Quantum error correction in scrambling dynamics and
  measurement-induced phase transition}.
\newblock \emph{\bibinfo{journal}{Phys. Rev. Lett.}}
  \textbf{\bibinfo{volume}{125}}, \bibinfo{pages}{030505}
  (\bibinfo{year}{2020}).
\newblock
  \urlprefix\url{https://link.aps.org/doi/10.1103/PhysRevLett.125.030505}.

\bibitem{Li19}
\bibinfo{author}{Li, Y.}, \bibinfo{author}{Chen, X.} \&
  \bibinfo{author}{Fisher, M. P.~A.}
\newblock \bibinfo{title}{Measurement-driven entanglement transition in hybrid
  quantum circuits}.
\newblock \emph{\bibinfo{journal}{Phys. Rev. B}}
  \textbf{\bibinfo{volume}{100}}, \bibinfo{pages}{134306}
  (\bibinfo{year}{2019}).
\newblock \urlprefix\url{https://link.aps.org/doi/10.1103/PhysRevB.100.134306}.

\bibitem{Jian20}
\bibinfo{author}{Jian, C.-M.}, \bibinfo{author}{You, Y.-Z.},
  \bibinfo{author}{Vasseur, R.} \& \bibinfo{author}{Ludwig, A. W.~W.}
\newblock \bibinfo{title}{Measurement-induced criticality in random quantum
  circuits}.
\newblock \emph{\bibinfo{journal}{Phys. Rev. B}}
  \textbf{\bibinfo{volume}{101}}, \bibinfo{pages}{104302}
  (\bibinfo{year}{2020}).
\newblock \urlprefix\url{https://link.aps.org/doi/10.1103/PhysRevB.101.104302}.

\bibitem{Bao20}
\bibinfo{author}{Bao, Y.}, \bibinfo{author}{Choi, S.} \&
  \bibinfo{author}{Altman, E.}
\newblock \bibinfo{title}{Theory of the phase transition in random unitary
  circuits with measurements}.
\newblock \emph{\bibinfo{journal}{Phys. Rev. B}}
  \textbf{\bibinfo{volume}{101}}, \bibinfo{pages}{104301}
  (\bibinfo{year}{2020}).
\newblock \urlprefix\url{https://link.aps.org/doi/10.1103/PhysRevB.101.104301}.

\bibitem{Aharonov00}
\bibinfo{author}{Aharonov, D.}
\newblock \bibinfo{title}{{Quantum to classical phase transition in noisy
  quantum computers}}.
\newblock \emph{\bibinfo{journal}{Phys. Rev. A}} \textbf{\bibinfo{volume}{62}},
  \bibinfo{pages}{062311} (\bibinfo{year}{2000}).

\bibitem{Gullans20b}
\bibinfo{author}{Gullans, M.~J.} \& \bibinfo{author}{Huse, D.~A.}
\newblock \bibinfo{title}{Scalable probes of measurement-induced criticality}.
\newblock \emph{\bibinfo{journal}{Phys. Rev. Lett.}}
  \textbf{\bibinfo{volume}{125}}, \bibinfo{pages}{070606}
  (\bibinfo{year}{2020}).
\newblock
  \urlprefix\url{https://link.aps.org/doi/10.1103/PhysRevLett.125.070606}.

\bibitem{Brydges2019}
\bibinfo{author}{Brydges, T.} \emph{et~al.}
\newblock \bibinfo{title}{Probing r{\'e}nyi entanglement entropy via randomized
  measurements}.
\newblock \emph{\bibinfo{journal}{Science}} \textbf{\bibinfo{volume}{364}},
  \bibinfo{pages}{260--263} (\bibinfo{year}{2019}).
\newblock \urlprefix\url{https://science.sciencemag.org/content/364/6437/260}.
\newblock
  \eprint{https://science.sciencemag.org/content/364/6437/260.full.pdf}.

\bibitem{Schrodinger36}
\bibinfo{author}{Schr{\"o}dinger, E.}
\newblock \bibinfo{title}{{Probability relations between separated systems}}.
\newblock \emph{\bibinfo{journal}{Math. Proc. Camb. Philos. Soc.}}
  \textbf{\bibinfo{volume}{32}}, \bibinfo{pages}{446--452}
  (\bibinfo{year}{1936}).

\bibitem{WinelandNobelRMP}
\bibinfo{author}{Wineland, D.~J.}
\newblock \bibinfo{title}{Nobel lecture: Superposition, entanglement, and
  raising schr\"odinger's cat}.
\newblock \emph{\bibinfo{journal}{Rev. Mod. Phys.}}
  \textbf{\bibinfo{volume}{85}}, \bibinfo{pages}{1103--1114}
  (\bibinfo{year}{2013}).
\newblock \urlprefix\url{https://link.aps.org/doi/10.1103/RevModPhys.85.1103}.

\bibitem{HarocheNobelRMP}
\bibinfo{author}{Haroche, S.}
\newblock \bibinfo{title}{Nobel lecture: Controlling photons in a box and
  exploring the quantum to classical boundary}.
\newblock \emph{\bibinfo{journal}{Rev. Mod. Phys.}}
  \textbf{\bibinfo{volume}{85}}, \bibinfo{pages}{1083--1102}
  (\bibinfo{year}{2013}).
\newblock \urlprefix\url{https://link.aps.org/doi/10.1103/RevModPhys.85.1083}.

\bibitem{jump2019}
\bibinfo{author}{Minev, Z.~K.} \emph{et~al.}
\newblock \bibinfo{title}{To catch and reverse a quantum jump mid-flight}.
\newblock \emph{\bibinfo{journal}{Nature}} \textbf{\bibinfo{volume}{570}},
  \bibinfo{pages}{200--204} (\bibinfo{year}{2019}).
\newblock \urlprefix\url{https://doi.org/10.1038/s41586-019-1287-z}.

\bibitem{Barreiro11}
\bibinfo{author}{Barreiro, J.~T.} \emph{et~al.}
\newblock \bibinfo{title}{{An open-system quantum simulator with trapped
  ions}}.
\newblock \emph{\bibinfo{journal}{Nature}} \textbf{\bibinfo{volume}{470}},
  \bibinfo{pages}{486--491} (\bibinfo{year}{2011}).

\bibitem{Yang20}
\bibinfo{author}{Yang, D.}, \bibinfo{author}{Grankin, A.},
  \bibinfo{author}{Sieberer, L.~M.}, \bibinfo{author}{Vasilyev, D.~V.} \&
  \bibinfo{author}{Zoller, P.}
\newblock \bibinfo{title}{{Quantum non-demolition measurement of a many-body
  Hamiltonian}}.
\newblock \emph{\bibinfo{journal}{Nature Commun.}}
  \textbf{\bibinfo{volume}{11}}, \bibinfo{pages}{1--8} (\bibinfo{year}{2020}).

\bibitem{vitale2021symmetryresolved}
\bibinfo{author}{Vitale, V.} \emph{et~al.}
\newblock \bibinfo{title}{Symmetry-resolved dynamical purification in synthetic
  quantum matter}.
\newblock \emph{\bibinfo{journal}{arXiv:2101.07814}}  (\bibinfo{year}{2021}).

\bibitem{Maunz2016}
\bibinfo{author}{Maunz, P. L.~W.}
\newblock \bibinfo{title}{High optical access trap 2.0.}
\newblock \emph{\bibinfo{journal}{Sandia National Laboratories Report No.
  SAND2016-0796R}}  (\bibinfo{year}{2016}).
\newblock
  \urlprefix\url{http://prod.sandia.gov/techlib/access-control.cgi/2016/160796r.pdf}.

\bibitem{egan2021faulttolerant}
\bibinfo{author}{Egan, L.} \emph{et~al.}
\newblock \bibinfo{title}{Fault-tolerant operation of a quantum
  error-correction code}.
\newblock \emph{\bibinfo{journal}{2009.11482}}  (\bibinfo{year}{2021}).

\bibitem{fossfeig2021entanglement}
\bibinfo{author}{Foss-Feig, M.} \emph{et~al.}
\newblock \bibinfo{title}{Entanglement from tensor networks on a trapped-ion
  qccd quantum computer}.
\newblock \emph{\bibinfo{journal}{arXiv:2104.11235}}  (\bibinfo{year}{2021}).

\bibitem{Gottesman98}
\bibinfo{author}{Gottesman, D.}
\newblock \bibinfo{title}{The heisenberg representation of quantum computers}.
\newblock In \emph{\bibinfo{booktitle}{Proc. XXII International Colloquium on
  Group Theoretical Methods in Physics}}, \bibinfo{pages}{32--43}
  (\bibinfo{publisher}{International Press}, \bibinfo{address}{Cambridge, MA},
  \bibinfo{year}{1998}).
\newblock \eprint{arXiv:quant-ph/9807006}.

\bibitem{Aaronson04}
\bibinfo{author}{Aaronson, S.} \& \bibinfo{author}{Gottesman, D.}
\newblock \bibinfo{title}{Improved simulation of stabilizer circuits}.
\newblock \emph{\bibinfo{journal}{Phys. Rev. A}} \textbf{\bibinfo{volume}{70}},
  \bibinfo{pages}{052328} (\bibinfo{year}{2004}).

\bibitem{Ippoliti21}
\bibinfo{author}{Ippoliti, M.}, \bibinfo{author}{Gullans, M.~J.},
  \bibinfo{author}{Gopalakrishnan, S.}, \bibinfo{author}{Huse, D.~A.} \&
  \bibinfo{author}{Khemani, V.}
\newblock \bibinfo{title}{Entanglement phase transitions in measurement-only
  dynamics}.
\newblock \emph{\bibinfo{journal}{Phys. Rev. X}} \textbf{\bibinfo{volume}{11}},
  \bibinfo{pages}{011030} (\bibinfo{year}{2021}).
\newblock \urlprefix\url{https://link.aps.org/doi/10.1103/PhysRevX.11.011030}.

\bibitem{cetina2020quantum}
\bibinfo{author}{Cetina, M.} \emph{et~al.}
\newblock \bibinfo{title}{Quantum gates on individually-addressed atomic qubits
  subject to noisy transverse motion}.
\newblock \emph{\bibinfo{journal}{arXiv:2007.06768}}  (\bibinfo{year}{2020}).

\bibitem{Lavasani21}
\bibinfo{author}{Lavasani, A.}, \bibinfo{author}{Alavirad, Y.} \&
  \bibinfo{author}{Barkeshli, M.}
\newblock \bibinfo{title}{{Measurement-induced topological entanglement
  transitions in symmetric random quantum circuits}}.
\newblock \emph{\bibinfo{journal}{Nature Phys.}} \textbf{\bibinfo{volume}{17}},
  \bibinfo{pages}{342--347} (\bibinfo{year}{2021}).

\bibitem{Sang20}
\bibinfo{author}{Sang, S.} \& \bibinfo{author}{Hsieh, T.~H.}
\newblock \bibinfo{title}{{Measurement Protected Quantum Phases}}.
\newblock \emph{\bibinfo{journal}{arXiv:2004.09509}}  (\bibinfo{year}{2020}).

\bibitem{Napp19}
\bibinfo{author}{Napp, J.}, \bibinfo{author}{La~Placa, R.~L.},
  \bibinfo{author}{Dalzell, A.~M.}, \bibinfo{author}{Brandao, F. G. S.~L.} \&
  \bibinfo{author}{Harrow, A.~W.}
\newblock \bibinfo{title}{{Efficient classical simulation of random shallow 2D
  quantum circuits}}.
\newblock \emph{\bibinfo{journal}{arXiv:2001.00021}}  (\bibinfo{year}{2020}).

\bibitem{Gullans20c}
\bibinfo{author}{Gullans, M.~J.}, \bibinfo{author}{Krastanov, S.},
  \bibinfo{author}{Huse, D.~A.}, \bibinfo{author}{Jiang, L.} \&
  \bibinfo{author}{Flammia, S.~T.}
\newblock \bibinfo{title}{{Quantum coding with low-depth random circuits}}.
\newblock \emph{\bibinfo{journal}{arXiv:2010.09775}}  (\bibinfo{year}{2020}).

\bibitem{maslov2017basic}
\bibinfo{author}{Maslov, D.}
\newblock \bibinfo{title}{Basic circuit compilation techniques for an ion-trap
  quantum machine}.
\newblock \emph{\bibinfo{journal}{New Journal of Physics}}
  \textbf{\bibinfo{volume}{19}}, \bibinfo{pages}{023035}
  (\bibinfo{year}{2017}).

\end{thebibliography}

\bibliographystyle{naturemag}

\section*{Acknowledgments}
We acknowledge fruitful discussions with E. Altman, S. Choi, A. Deshpande, S. Diehl, B. Fefferman, S. Gopalakrishnan, M. Ippoliti, V. Khemani, A. Nahum, J. Pixley, O. Shtanko, and A. Zabalo  and the contributions of M. Goldman, K. Beck, J. Amini, K. Hudek, and J. Mizrahi to the experimental setup. 
\textbf{Funding}
This work is supported by the ARO through the IARPA LogiQ program, the NSF STAQ Program, the AFOSR MURIs on Dissipation Engineering in Open Quantum Systems and Quantum Measurement/Verification and Quantum Interactive Protocols, the ARO MURI on Modular Quantum Circuits, the DoE Quantum Systems Accelerator, the DoE ASCR Accelerated Research in
Quantum Computing program (award No. DE-SC0020312). L. Egan is also funded by NSF award DMR-1747426.
This work was performed at the University of Maryland with no material support from IonQ.
\textbf{Author contributions}
C.N. collected the data. C.N. and P.N analyzed the data. C.N., P.N., and M.J.G. wrote the manuscript and designed figures. M.C. and C.M. led construction of the experimental apparatus with contributions from  C.N., D.Z., A.R., L.E., and D.B. Theory support was provided by P.N., M.J.G., A.V.G. and D.A.H.. C.M., M.J.G., and D.A.H. supervised the project. All authors discussed results and contributed to the manuscript.
\textbf{Competing interests}
No authors claim a competing interest.
\textbf{Data and Materials}
All data is available in the manuscript or the Supplementary Materials.

\section*{List of Supplementary Materials}

Materials and Methods
\\
Supplementary Text
\\
Table S1
\\
Fig S1-S6

\clearpage
\pagenumbering{arabic}
\section*{Supplementary Materials}

\subsection*{Materials and Methods}

\paragraph*{Scrambling Unitary} A scrambling unitary, $U_S$, is applied after the system is entangled with the reference, before the random time evolution begins. The scrambling unitary consists of $4$ layers: odd-numbered layers are composed of single-qubit operations on each qubit and even-numbered layers are composed of fully entangling XX($\pi/4$) gates on $L/2$ random qubit-pairs.

\renewcommand{\thefigure}{S\arabic{figure}}

\setcounter{figure}{0}

\begin{figure}[!ht]
\begin{centering}
\includegraphics[scale=.88]{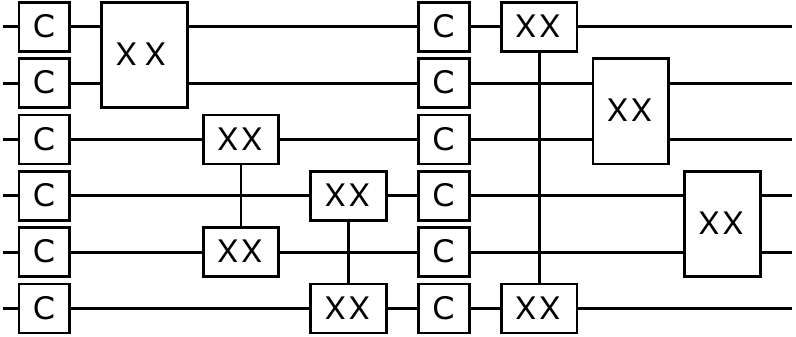}
\caption{\label{fig:scrambling-unitary} Example of a scrambling unitary on a system with $L=6$ qubits. Each single-qubit gate $C$ refers to a random single-qubit Clifford gate. The XX gates have an implied rotation angle of $\pi/4$. }
\end{centering}
\end{figure}

\paragraph*{Measurement Protocol}
In our circuit ensemble, each gate after the scrambling layer is followed by a probabilistic measurement. Given the constraints of the hardware, we choose a measurement strategy that reduces the number of measurements. In addition, the ensemble generated with our measurement strategy scales to system sizes that are beyond the reach of available hardware and can only be studied with numerical simulations. 

We maintain a list $\mathcal{M}$ which is initialized to all system qubits in the beginning of the circuit. After each gate, we measure one of the qubits involved in the gate with probability $p$. Having decided to perform a measurement after a XX gate, we randomly choose the qubit to measure and the basis of measurement.  If both qubits participating in the XX gate are in $\mathcal{M}$, we randomly select one with probability half and measure it in $X$ basis with probability $p_x$ and in $Z$ basis with probability $1-p_x$. If only one of the qubits is in $\mathcal{M}$, we measure that qubit (in the $X$ basis with probability $p_x$ and in the Z basis with probability ($1-p_x$). If neither qubit is in $\mathcal{M}$, we do not measure any. Measurement outcomes in Clifford circuits are deterministic or are equally likely to be zero or one. In the absence of noise, measuring a qubit with a deterministic outcome has no effect on the purification of the reference.  As a result, we only measure qubits with non-deterministic outcomes. Additionally, after each physically performed measurement, we remove the measured qubit from $\mathcal{M}$. Once $|\mathcal{M}| = L-4$, we reinitialize the list with all the qubits in the system. With a low measurement probability, $p=0.15$, used in our experiment the number of measurements in the circuits investigated are less than 4, and thus the the list $\mathcal{M}$ need not be reinitialized. This ensures that no system qubit in the experiment is measured more than once. 

\paragraph*{Feedback} 
The feedback circuit is added at the end to disentangle the reference from the ancillae qubits. In the pure phase, the reference qubit purifies in one of $x$, $y$ or $z$ bases and its state (0 or 1) depends on the projections induced upon the measurement ancillae. The basis of purification can be anticipated with classical simulation of the Clifford circuit. A single-qubit rotation is performed on the reference qubit to ensure that it returns to the $z$-basis following purification. Since we do not have access to the measurement outcome until the very end of the circuit, we construct a logic circuit, consisting of CNOT gates, to ensure that the reference qubit purifies to the zero state. This is done by classically anticipating the entanglement between measurement ancillae and reference qubit, then generating a sequence of CNOT gates to disentangle the reference.

For example, in the batch $L=4, p_x=0$, circuit $\#45$ purifies the reference in the $x$-basis. There are three measurements. The outcomes of the measurement ancillae and the reference qubit are related by the truth table in Table \ref{tab:measurement-reference}.

\renewcommand{\thetable}{S\arabic{table}}

\setcounter{table}{0}

\begin{table}[!ht]
    \centering
\begin{tabular}{|c|c|}
\hline
    Measurement Record & Reference State \\
    \hline
    001 & \multirow{4}{*}{0} \\ 
    010 & \\
    101 & \\
    110 & \\
    \hline
    000 & \multirow{4}{*}{1} \\ 
    011 & \\
    100 & \\
    111 & \\
    \hline
\end{tabular}
    \caption{\textbf{Feedback Truth Table} Truth table for outcomes of measurement ancillae and reference qubit for a circuit.}
    \label{tab:measurement-reference}
\end{table}

\begin{figure}[!ht]
    \centering
    \includegraphics[width=0.4\textwidth]{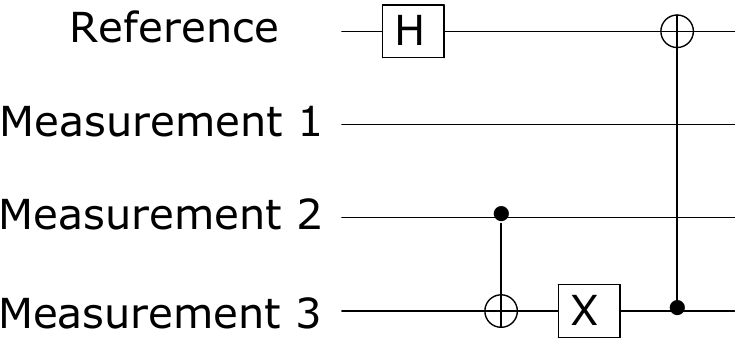}
    \label{circuit:feedback}
\end{figure}

The feedback circuit, in this particular case, is given by the circuit diagram in \ref{circuit:feedback}. The Hadamard gate is used to align the reference along the $z$-basis. The following sequence of CNOT and $X$ gates implement the logic to disentangle the reference from the ancillae. When implementing the circuit, all CNOT gates are compiled to XX gates \cite{maslov2017basic}.

\paragraph*{Circuit Optimization} For each circuit, commuting single-qubit rotations and $XX(\pi/4)$ gates are merged, wherever possible, to reduce the size of the circuit.

\begin{figure}[h]
\begin{centering}
\includegraphics[scale=0.8]{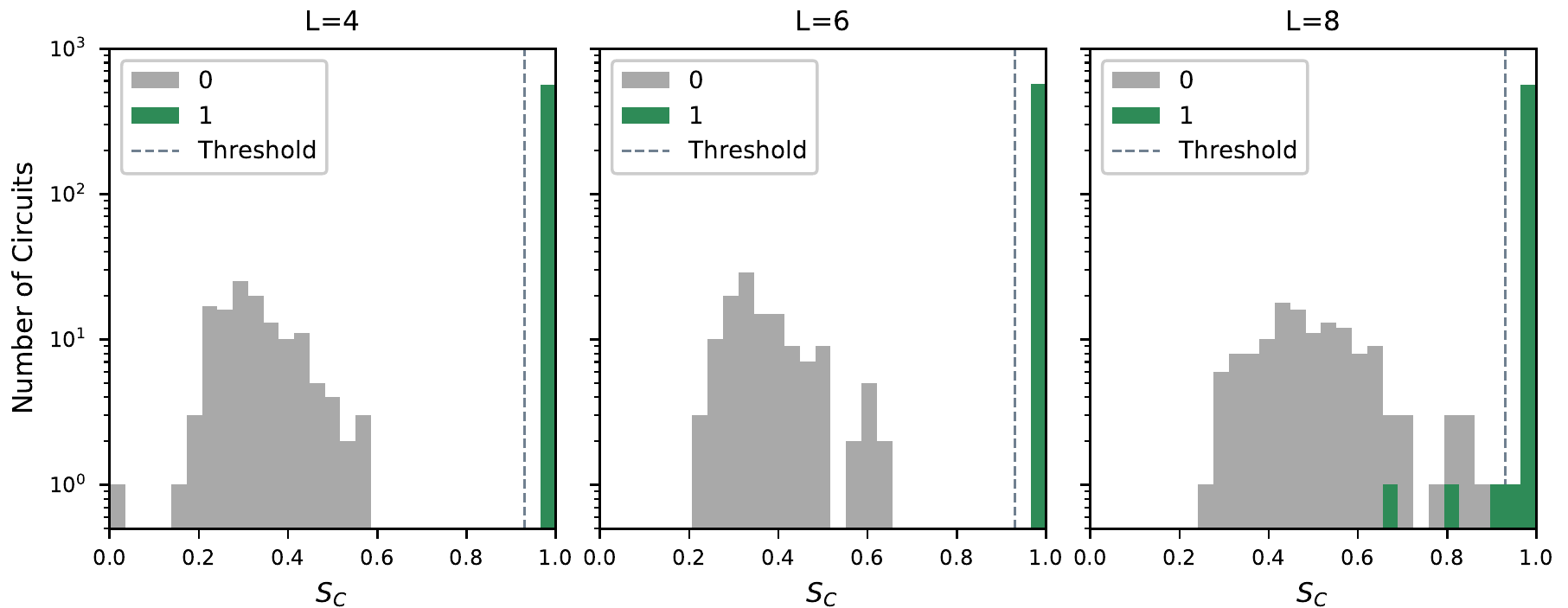}
\caption{\label{fig:supp_histograms} \textbf{Histogram of Experimental Data for $S_C$} All raw outcomes of $S_C$ in study of phases (main text Fig.~3A). The legend indicates the simulated expected outcome for that circuit. The bin size is .033 and $S_C=.93$ (dashed line) is used as a threshold for all the data.}
\end{centering}
\end{figure}

\begin{figure}[h]
\begin{centering}
\includegraphics[scale=0.9]{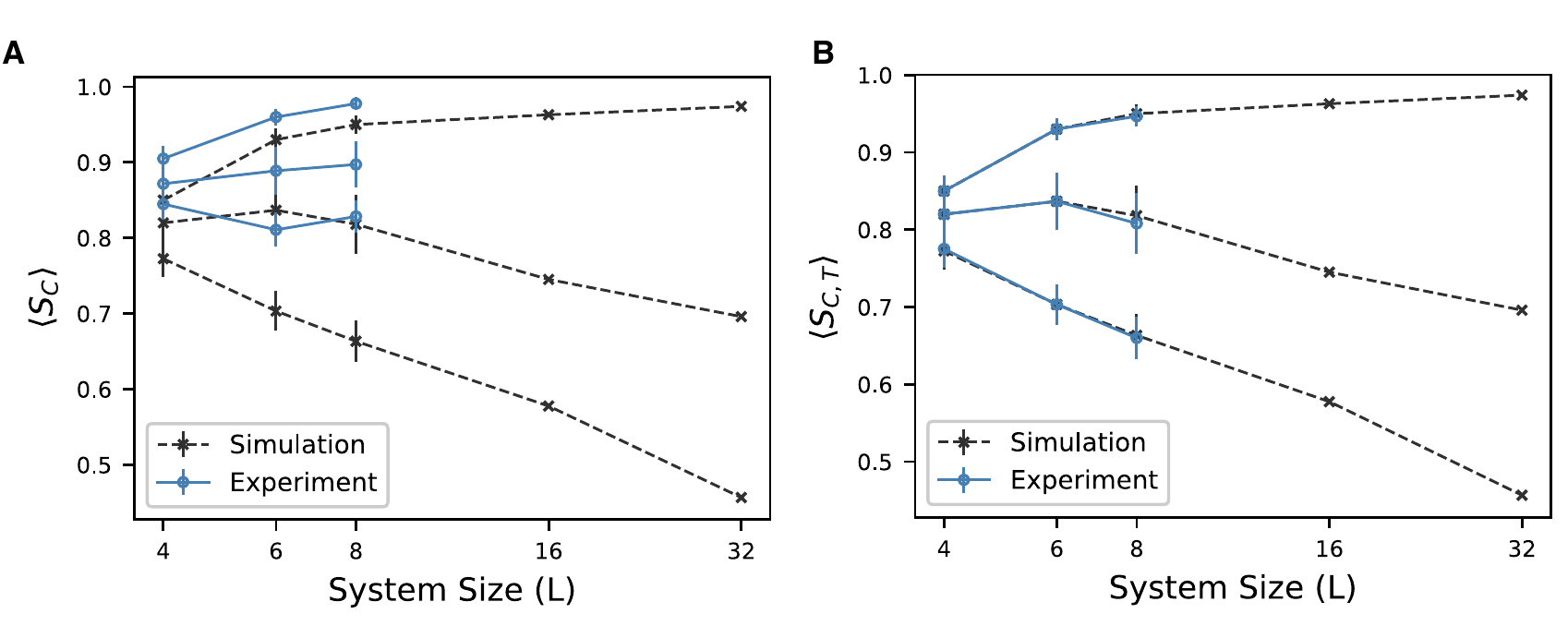}
\caption{\label{fig:supp_rawdata} \textbf{Comparison of Theory and Experiment} (A) Raw average of all circuit outcomes without thresholding applied. (B) Thresholded data with extended simulations showing expected behavior up to L=32.}
\end{centering}
\end{figure}

\paragraph*{Raw Data and Thresholding}
Data presented in the main text Fig.~3A is presented after binning via a threshold. Fig.~\ref{fig:supp_histograms} shows histograms of the outcomes for all circuits with each entropy averaged over the outcome of 1000 shots per circuit. Furthermore, Fig.~\ref{fig:supp_rawdata}A shows the average classical entropy over all circuits for each system size and $p_x$ value. These average are clearly much higher than simulation. 

The primary reason for the discrepancy between the simulations and the experimental data is that the simulations do not include noise.  When including realistic noise sources in the simulations (see Noise Model section in Supplementary Text), such as dephasing or gate amplitude errors, we find that the entropy outcomes are pushed to higher values. Nevertheless, we see in the data for L=4 and L = 6 that there is a clear separation observed in the entropy values for circuits which are mixed or pure (Fig. S2). This separation is also evident in Fig. 1B, with the clear jump from mixed to pure in the evolution of the circuit. 

We can use these properties of the circuits to mitigate  noise effects. In the final data processing, we assume a Gaussian distribution of expected $S_C=0$ circuit outcomes and $S_C=1$ circuit outcomes and find their intersection, which is used as a threshold at $S_C=0.93$. Circuit outcomes below the threshold are counted as 0 and outcomes above are counted as 1. We find three thresholded circuit outcomes disagree with the simulated expected value for that circuit, for an error of 3/699 circuits for the $L=8$ case. Fig.~\ref{fig:supp_rawdata}B shows the result after all processing alongside simulations of the exact circuits for sizes $L=(4,6,8)$ and representative samples for sizes $L=(16,32)$. The same threshold is used for all system sizes.

\clearpage

\paragraph*{Critical Scaling Theory}

\begin{figure}[htbp]
\begin{center}
\includegraphics[width=.9\textwidth]{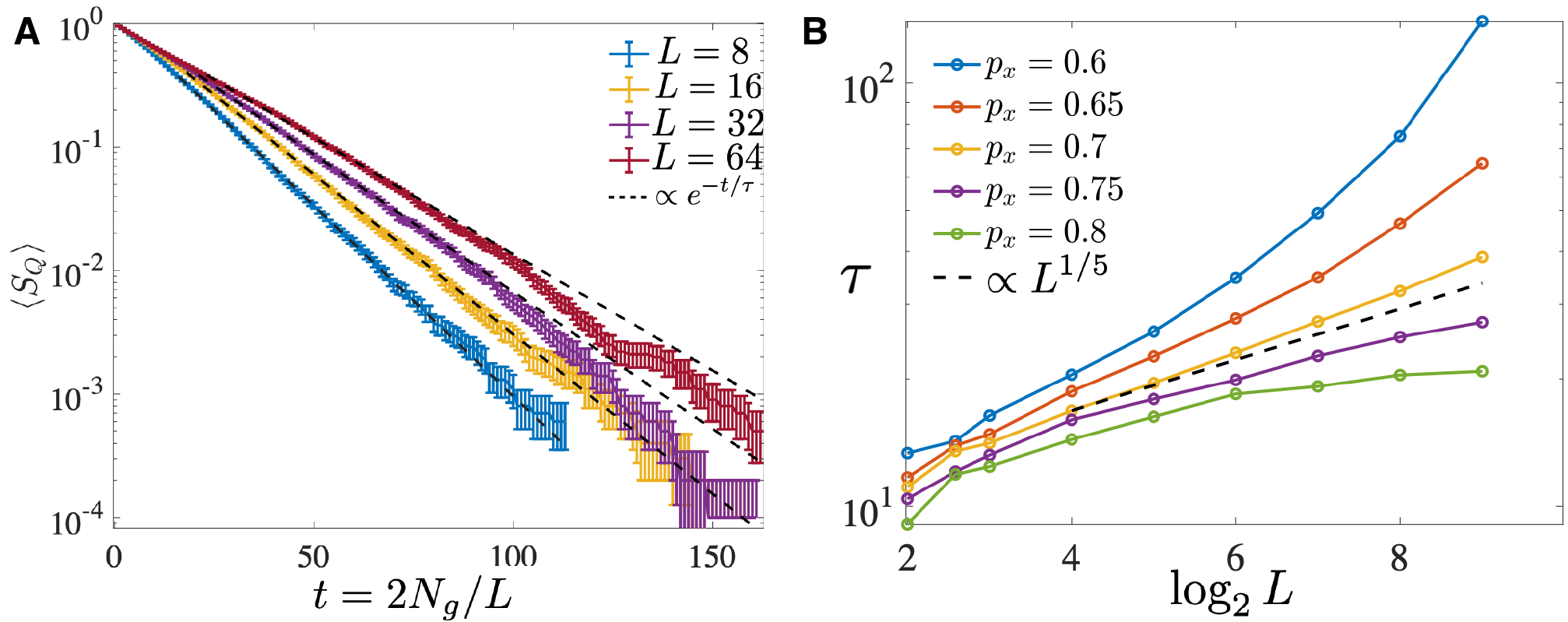}
\caption{\textbf{Analysis Method to Extract Critical Data} (a) Late time decay of $\langle S_Q \rangle$ showing the exponential decay regime used to extract the decay rate $\tau$.  Here, we took $(p,p_x) = (0.15,0.7)$ near the critical point.  (b) Scaling of $\tau$ vs $L$ for different values of $p_x$ at $p=0.15$.  We can estimate $p_{xc}$ and extract $z$ by looking for the inflection point in this family of curves and fitting the slope. }
\label{fig:decaytime}
\end{center}
\end{figure}

Our method for locating the critical point in these all-to-all models is illustrated in Fig.~\ref{fig:decaytime}.  For $p_x \sim p_{xc}$, we can run the dynamics out to a time where $\langle S_Q(t) \rangle$ exhibits a simple exponential decay $\propto e^{-t/\tau}$.  We then use least squares fitting to find the exponential decay rate $\tau$ for each value of $p_x$ and $L$.  
Deep in the mixed phase, $\tau$ diverges exponentially with $L$ \cite{Gullans20}, while in the pure phase $\tau$ approaches a constant independent of system size. At the critical point ($p= p_{xc}$), $\tau \sim L^{z}$, where $z$ is the dynamical critical exponent.  Thus, we can estimate $p_{xc}$ by looking for the value of $p_x$ where $\tau(L)$ goes through an inflection point on a log-log plot.  This behavior is illustrated in Fig.~\ref{fig:decaytime}(b) for the model with $p= 0.15$ and $|\mathcal{M}| \ge L-4)$.  Near $p_x = 0.7-0.75$, we see that the decay rate $\tau$ grows as power law $L^{1/5}$ over the given range of sizes.  This value of $z=1/5$ is consistent with the scaling one would expect from mean-field percolation.  The close ties between these phase transitions and percolation  have been noted in past works.  Notably, for the Hartley entropy of Haar random circuits with measurements, there is an exact mapping to a percolation problem in the circuit geometry \cite{Skinner19}.  In the all-to-all setting considered here, this mapping also predicts $z=1/5$.

Using this estimate for $z$, we can accurately measure the critical point $p_{xc}$ and critical exponent $\nu$ of the purification transition using the method illustrated in Fig.~\ref{fig:results}B. 
We hypothesize a scaling form for $\tau$
\begin{equation}
\tau = L^{z} f[ (p_x - p_{xc})L^{z/\nu}],
\end{equation}
which predicts that a crossing will appear with increasing sizes when plotting $\tau/L^{z}$ vs $p_x$.  We see consistent results with this scaling assumption in Fig.~\ref{fig:results}B, from which we locate $p_{xc} = 0.72(1)$.   A similar analysis was used for other values of $p \ne 0.15$ to extract the phase diagram in Fig.~\ref{fig:diagram}A.
 After locating $p_{xc}$, we then collapse the data as shown in the inset to Fig.~\ref{fig:results}B to obtain an estimate $\nu = 1/2$, which is also consistent with the predication from mean-field percolation.  We leave a more detailed analysis of the critical properties of this model for future work.

\subsection*{Supplementary Text}

\begin{figure}[h]
\begin{centering}
\includegraphics[scale=0.85]{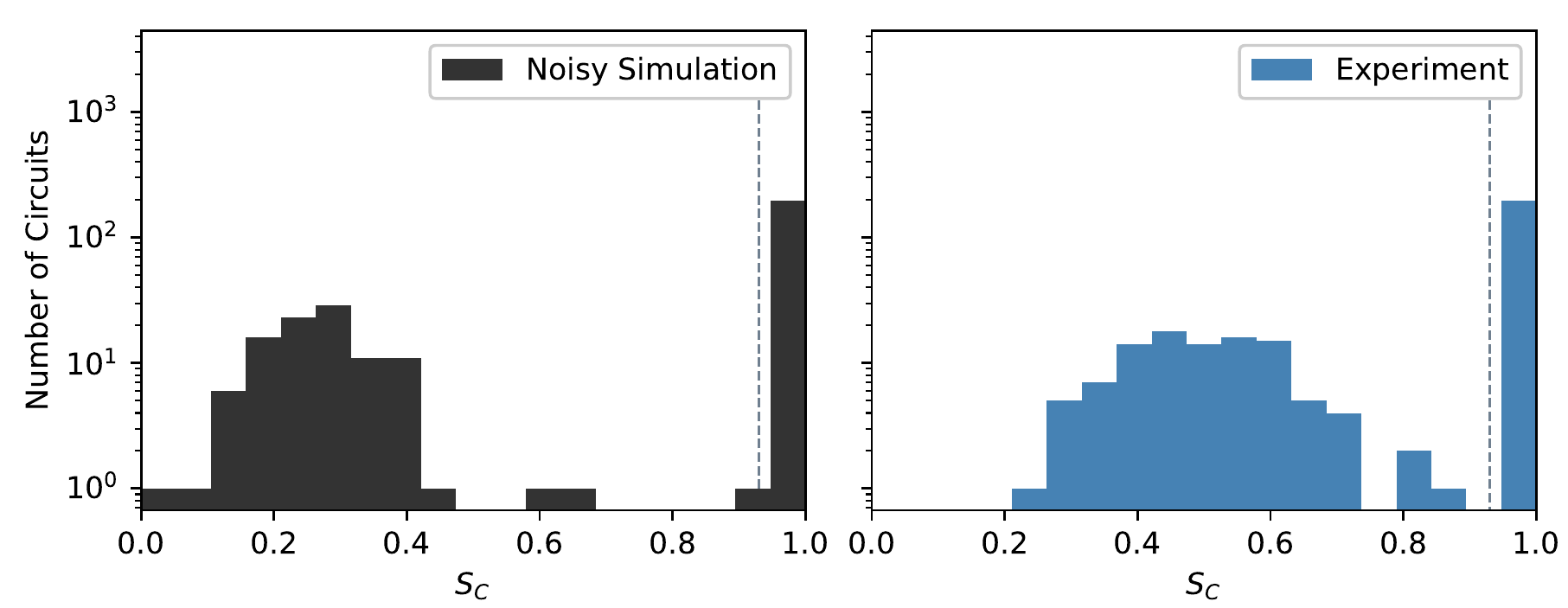}
\caption{\textbf{Comparison of Noisy Simulations to Experiment} Example results of a noisy simulation for all circuits corresponding to $L=8$ and $p_x=1$ (left) compared to the experimental outcomes (right). Bin size is .05. Dashed lines show the threshold of $S_C=0.93$ \label{fig:noisy_sim}}
\end{centering}
\end{figure}

\paragraph*{Noise Model}
For the noisy simulation, we assume a simple model of XX-gate crosstalk, which is the dominant error mechanism in this work. The crosstalk value is predicted from the measured single-qubit Rabi crosstalk and the participation matrix for each gate \cite{egan2021faulttolerant}. With an increase of 50\% above the predicted crosstalk value, we find that the noisy simulation is qualitatively similar to the experimental results. Such an increase could easily be caused by a shift in the ion positions relative to the individual-addressing beams, since the addressing crosstalk increases sharply for small ion displacements from their optimal positions. A sample noisy simulation is shown in Fig.~\ref{fig:noisy_sim} for the case of $L=8$ and $p_x=1$. Any additional noise that decoheres the state of the reference or system will further shift the distribution of outcome entropies upwards, towards the mixed state. Other noise mechanisms that likely contribute to the shift observed in the data include $T_2$ dephasing, random over/under rotation errors caused by beam position fluctuations, spurious entanglement of qubits with the axial modes of the ion chain, and SPAM errors.

\paragraph*{Error Mitigation}

\begin{figure}[h]
\begin{centering}
\includegraphics[scale=0.9]{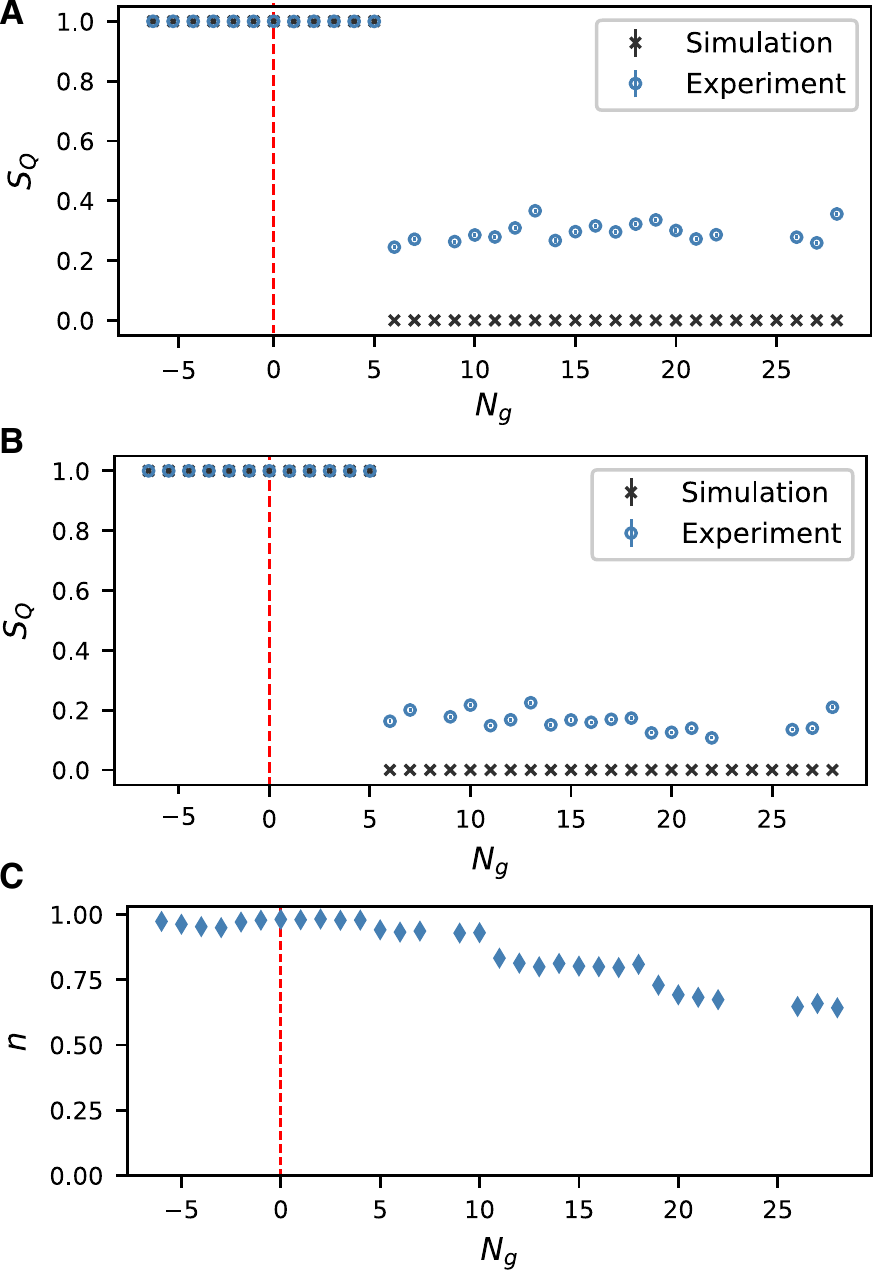}
\caption{\textbf{Error Mitigation on  Purifying Circuit} Time evolution of a sample $L=6$ circuit, conditioned on a particular choice of outcomes for the intermediate measurements, that purifies without (A) and with (B) error mitigation applied. (C) Average proportion of shots retained, $n$,  for each time step \label{fig:error-mitigated}}
\end{centering}
\end{figure}

An interesting aspect of noiseless stabilizer circuits is that measurements of Pauli operators often have deterministic expectation values.  We can use this fact to aid error mitigation of the noisy implementation of these circuits. As we scale the system to larger sizes or higher circuit depths, this type of error mitigation may become useful. Here, we describe those strategies and show data from Fig.~1B without and with these error mitigation techniques (Fig.~\ref{fig:error-mitigated}A-B).  For simplicity, all the data presented in the main text does not use these error mitigation techniques.

In the ideal implementation of the circuit, certain qubits have a deterministic output in $x$, $y$ or $z$ basis at the end of a circuit. In addition, the qubits that do not participate in the purification dynamics should ideally be in the zero state in the computational ($z$-basis) state. The basis and the respective deterministic outcome can be anticipated with classical simulation. If a qubit is deterministic along $x$ or $y$ basis, we append a single-qubit rotation to align that qubit along the $z$-basis. In the error mitigation stage, we discard the records where the non-participating qubits read a value other than zero or the deterministic qubits, which now should all be purified in the $z$-basis, do not match the simulated expectation.  Note, that this method relies on post-selection, so is not directly scalable to the large-$L$ limit.  Eventually, active error correction, or similar techniques, would need to be applied to design scalable protocols to probe the ideal circuit evolution in the presence of  noise. 

To investigate the amount of data discarded by error-mitigation, we consider the quantity $n = \langle \left(\sum_{i} N_{b, i}\right)/N \rangle_{b \in \{x, y, z\}}$, which measures the proportion of observations retained  after error-mitigation, aggregated by measurement outcomes and averaged across the three bases. Here $N_{b, i}$ is the number of observations for a tomography circuit used to measure the Pauli expectation along basis $b$, that, conditioned on the measurement record reading $i$, have deterministic qubits matching simulation, and $N$ is the total shots for each circuit used in tomography, which in this instance is 4000. These data were taken in a random order. In (Fig.~\ref{fig:error-mitigated}C), we plot the $n$ for each time-step of (Fig.~\ref{fig:error-mitigated}B). As expected, the proportion of erroneous observations increases with larger circuit depths.




\end{document}